\documentclass[conference]{IEEEtran}
\IEEEoverridecommandlockouts
\usepackage{cite}
\usepackage{amsmath,amssymb,amsfonts}
\usepackage{algorithmic}
\usepackage{textcomp}

\usepackage{enumitem}
\usepackage{multirow}
\usepackage{booktabs}
\usepackage{array}
\usepackage{multirow}
\usepackage{balance}
\usepackage{hyperref}
\usepackage{url}

\usepackage{graphicx}
\usepackage[export]{adjustbox} 
\usepackage{xcolor}
\usepackage{array}
\usepackage{booktabs}
\usepackage{multirow}
\usepackage[inkscapelatex=false]{svg}
\newcolumntype{H}{>{\setbox0=\hbox\bgroup}c<{\egroup}@{}}

\newcommand{\ours}{Music4All~A+A}

\def\BibTeX{{\rm B\kern-.05em{\sc i\kern-.025em b}\kern-.08em
    T\kern-.1667em\lower.7ex\hbox{E}\kern-.125emX}}
\begin{document}

\title{Music4All A+A: A Multimodal Dataset for Music Information Retrieval Tasks}

\author{Anonymous Author(s)}

\author{Jonas Geiger$^{1}$, Marta Moscati$^{1}$, Shah Nawaz$^{1}$, Markus Schedl$^{1,2}$  \\
$^{1}$Johannes Kepler University Linz, Austria, \\
$^{2}$Human-centered AI Group, AI Lab, Linz Institute of Technology, Austria \\
\tt jonasgeiger@outlook.de, \{marta.moscati, shah.nawaz, markus.schedl\}@jku.at
}


\maketitle

\begin{abstract}
    Music is characterized by aspects related to different modalities, such as the audio signal, the lyrics, or the music video clips. This has motivated the development of multimodal datasets and methods for Music Information Retrieval (MIR) tasks such as genre classification or autotagging. Music can be described at different levels of granularity, for instance defining genres at the level of artists or music albums. However, most datasets for multimodal MIR neglect this aspect and provide data at the level of individual music tracks. We aim to fill this gap by providing Music4All Artist and Album (\ours), a dataset for multimodal MIR tasks based on music \textit{artists} and \textit{albums}. \ours{}  is built on top of the Music4All-Onion dataset, an existing track-level dataset for MIR tasks. \ours{}  provides metadata, genre labels, image representations, and textual descriptors for $6{,}741$ artists and $19{,}511$ albums. Furthermore, since \ours{}  is built on top of Music4All-Onion, it allows access to other multimodal data at the track level, including user--item interaction data. This renders \ours{}  suitable for a broad range of MIR tasks, including multimodal music recommendation, at several levels of granularity. To showcase the use of \ours{} , we carry out experiments on multimodal genre classification of artists and albums, including an analysis in missing-modality scenarios, and a quantitative comparison with genre classification in the movie domain. Our experiments show that images are more informative for classifying the genres of artists and albums, and that several multimodal models for genre classification struggle in generalizing across domains. We provide the code to reproduce our experiments at \url{https://github.com/hcai-mms/Music4All-A-A}, the dataset will be linked in the repository at camera-ready and provided open-source under a CC BY-NC-SA 4.0 license. 
\end{abstract}

\begin{IEEEkeywords}
    Multimodal Learning, Vision and Language, Missing Modalities, Music Information Retrieval
\end{IEEEkeywords}

\section{Introduction and Motivation}
   Multimedia data is complex and involves information stemming from different modalities~\cite{baltruvsaitis2018multimodal,xu2023multimodal}. This is also the case for music, since data from different sources such as audio, lyrics, or music video clips are all relevant in the characterization of music tracks. Traditionally, music information retrieval (MIR) has focused mostly on the audio signal to address tasks such as music genre classification or autotagging~\cite{bello2008genre_classification,seyerlehner2010blf_genre_classification}. In the past decades, however, the advancements in multimodal learning have enabled the use of multimodal data also for MIR tasks~\cite{CLAP2022,zhao2023clap,laionclap2023,huang2022mulan,Koepke2022audio_retrieval,deldjoo2024content}. These advancements in multimodal MIR were enabled by the emergence of multimodal datasets related to music. Several multimodal music datasets have focused on the task of music genre classification~\cite{oramas2017multi,oguike2025multimodal}, and rely on the use of audio, textual, and visual representations of music tracks. This task is particularly relevant for music since genres are useful labels to identify tracks sharing similar semantics and to help 
   browse large music catalogs. 
   In the domain of music, due to the shorter consumption time of tracks compared to movies or books, for example, classifying items by genre might be more useful when done at a higher granularity than individual tracks. For instance, classifying the genre of an artist from a picture of one of their live shows 
   might facilitate tracking new releases of a specific genre. However, most existing datasets address the problem of music genre classification at the track level, with the exception of the MuMu dataset~\cite{oramas2017multi}, which considers albums. 
    In addition, to our knowledge, all existing datasets provide audio, textual, visual data, or combinations thereof, neglecting one aspect that is relevant in characterizing music on large-scale streaming platforms, such as their co-occurrence in the streaming profiles of users. This both limits the perspective provided on music, and prevents the use of multimodal data for the task of music recommendation. We aim to fill these gaps by providing Music4All Artist and Album (\ours), a dataset of multimodal representations of music \textit{artists} and \textit{albums}. Our dataset is built on top of the multimodal music dataset Music4All-Onion~\cite{moscati2022music4all_onion}, and this allows mapping our dataset to user--item interaction data as well. We extend the dataset with additional genre information, image representations, and textual descriptors for $6{,}741$ artists and $19{,}511$ albums. Compared to Music4All-Onion and other existing multimodal MIR datasets, ours offers a different granularity and a more balanced distribution of genre labels. 
    Furthermore, since it is built on top Music4All-Onion~\cite{moscati2022music4all_onion}, it can be extended with track-level representations of the audio, video clips, as well as user--item interaction data, which enables further uses, 
    e.g., for music recommendation tasks. 

    In this paper, we benchmark the \ours{}  dataset on the task of music genre classification from multimodal data, both at the artist and album level, and including missing-modality scenarios. Leveraging several well-established baselines for genre classification, our analysis shows that the image modality better represents music genres compared to textual descriptors, and that several baselines that are competitive in other domains, such as movie genre classification, struggle in generalizing to the domain of music, highlighting the need for more datasets for multimodal MIR, such as the proposed one.

\section{Related Work}
A comprehensive list of relevant publicly available datasets is provided in Table~\ref{tab:datasets}, including both those related to unimodal and multimodal music genre classification, and those related to multimodal genre classification beyond the music domain.  
\subsection{Music Genre Classification Datasets}
    Over the past few years, many datasets (e.g.,~\cite{oramas2017multi,arevalogated2017,oguike2025multimodal,moscati2022music4all_onion}) have been proposed to support the research on unimodal and multimodal music genre classification. 
    Among unimodal datasets, the GTZAN dataset~\cite{tzanetakis2002musical} stands as one of the most widely-used resources for music genre classification, serving as a benchmark in audio-based genre classification~\cite{sturm2012survey}.
    Similarly, the Extended Ballroom dataset~\cite{marchand2016extended} provides a useful resource for audio-based genre classification. Going beyond unimodal, audio-based datasets, Oramas et al.~\cite{oramas2017multi} developed the MuMu dataset, which comprises $31,000$ music album entries, each accompanied by cover image, text reviews, and audio tracks. Similarly, Oguike et al.~\cite{oguike2025multimodal} curated a dataset of $2,000$ music videos, each containing audio tracks, album covers, and lyrics, further expanding the resources available for multimodal genre analysis.
\subsection{Multimodal Genre Classification Datasets}
    In the multimodal domain, several large-scale datasets have emerged to support genre classification across diverse media types.
    For example, Iwana et al.~\cite{iwana2016judging} curated a dataset of $137,788$ books, incorporating cover images, titles, and author texts. 
    Similarly, Arevalo et al.~\cite{arevalogated2017} curated the Multimodal IMDb (MM-IMDb) dataset, comprising $25,959$ movie titles, corresponding plot summaries, images, and metadata. 
    More recently, Jiang et al.~\cite{jiang2023deep} expanded the landscape of multimodal genre classification by introducing a large-scale dataset of $50,000$ video games, incorporating cover artworks, descriptions, and titles. 

    While our work shares the multimodal approach of MM-IMDb~\cite{arevalogated2017} (movie genres) and MuMu~\cite{oramas2017multi} (music genres), it introduces several  advancements. Unlike MM-IMDb, we focus on the music domain with multi-label genre classification.
    More importantly, our dataset overcomes the class imbalance limitations of the MuMu dataset that biases models towards the most frequently occurring genres. Furthermore, we provide genres and multimodal data at two levels of granularity: albums and artists. Table~\ref{tab:top_10_genres_extended} compares the top $10$ most represented genres in MuMu, MM-IMDb, and the proposed \ours{} dataset. Finally, the connection with the Music4All-Onion dataset allows future work on music recommendation at different levels of granularity (music track, album, or artist).

\begin{table*}
\centering
  \caption{Publicly available datasets for genre classification.  
  }
  \begin{tabular}{llllcc}
    \toprule
    Dataset & Type     & Domain & Modalities & Size & Genre \\
    \toprule
    GTZAN~\cite{tzanetakis2002musical}   & Unimodal   & Music  & Audio                 & 1,000  & 10\\
    Extended Ballroom~\cite{marchand2016extended} & Unimodal   & Music  & Audio        & 4,180  & 13 \\
    Book Cover\cite{iwana2016judging}    & Multimodal   & Literature & Image \& Text   & 57,000 & 30 \\
    Jiang et al.~\cite{jiang2023deep}    & Multimodal & Games  &  Image \& Text & 50,000 & 15 \\
    Sotho-Tswana~\cite{oguike2025multimodal} &  Multimodal & Music & Audio, Text \& Image & $\sim$2,000 & 18 \\
    \midrule
    MuMu~\cite{oramas2017multi}          & Multimodal & Music  & Audio, Image, \& Text & 31,000 & 250 \\
    MM-IMDb~\cite{arevalogated2017}      & Multimodal & Movies & Image \& Text & 25,959 & 26  \\
    \midrule
    \ours{} (Artist) & Multimodal & Music & Image \& Text & 6{,}741 & 659 \\
    \ours{} (Album) & Multimodal & Music  & Image \& Text & 19{,}511 & 737 \\
    
    \bottomrule
  \end{tabular}
  \label{tab:datasets}

\end{table*}

\begin{table*}[htbp]
    \centering
    \caption{Top-10 most represented genres for Albums in MuMu~\cite{oramas2017multi}, MM-IMDb (ours), and our Music4All-derived dataset (Albums Ours, Artists Ours).}
    \label{tab:top_10_genres_extended}
    \resizebox{0.99\textwidth}{!}{
    \begin{tabular}{@{}lc@{\quad\quad}lc@{\quad\quad}lc@{\quad\quad}lc@{}}
        \toprule
        \multicolumn{2}{@{}c}{\textbf{Albums MuMu}~\cite{oramas2017multi}} & \multicolumn{2}{@{}c}{\textbf{MM-IMDb}\cite{arevalogated2017}} & \multicolumn{2}{@{}c}{\textbf{\ours{} (Album)}} & \multicolumn{2}{c}{\textbf{\ours{} (Artist) }} \\
        \cmidrule(r{0.5em}){1-2} \cmidrule(lr{0.5em}){3-4} \cmidrule(lr{0.5em}){5-6} \cmidrule(l{0.5em}){7-8}
        Genre & \% of Albums & Genre & \% of MM-IMDb & Genre & \% of Albums & Genre & \% of Artists \\
        \midrule
         Pop & 84.38 & Drama & 53.80 & Rock & 67.00 & Pop & 57.56  \\
         Rock & 55.29 & Comedy & 33.10 & Pop & 59.49 & Rock & 54.68  \\
         Alternative Rock & 27.69 & Romance & 20.66 & Alternative Rock & 39.76 & Soundtrack & 29.79 \\
         World Music & 19.31 & Thriller & 20.00 & Pop Rock & 31.14 & Alternative Rock & 29.02  \\
         Jazz & 14.73 & Crime & 14.78 & Singer Songwriter & 30.52 & Singer Songwriter & 26.85   \\
         Dance \& Electronic & 12.23 & Action & 13.68 & Indie Rock & 30.21 & Easy Listening & 26.02  \\
         Metal & 11.50 & Adventure & 10.44 & Classic Rock & 26.30 &  Pop Rock & 24.51 \\
         Indie \& Lo-Fi & 10.45 & Horror & 10.41 & Experimental & 25.50 & Classic Rock & 23.36   \\
         R\&B & 10.10 & Documentary & 8.02 & Easy Listening & 25.43 &  Indie Rock & 23.07 \\
         Folk & 9.69 & Mystery & 7.92 & Metal & 25.06 &  Soul & 22.15 \\
        \bottomrule
    \end{tabular}
    }
\end{table*}

\begin{table*}[htbp] 
\centering 
\caption{Examples of model predictions for \ours{} artist items. \textcolor{blue}{Blue} genres are true positives, \textcolor{red}{red} are false positives.} 
\label{tab:qualitative_sbnet_clip_artist_variants} 
\begin{tabular}{l c >{\raggedright\arraybackslash}p{5cm} >{\raggedright\arraybackslash}p{6cm} } 
\toprule 
\textbf{Artist} & \textbf{Image} & \textbf{Info / Model} & \textbf{Predicted / Ground Truth Genres} \\ 
\midrule 
\multirow{5}{*}{\textbf{Beast in Black}} & \multirow{5}{*}{\includegraphics[width=2cm,valign=c]{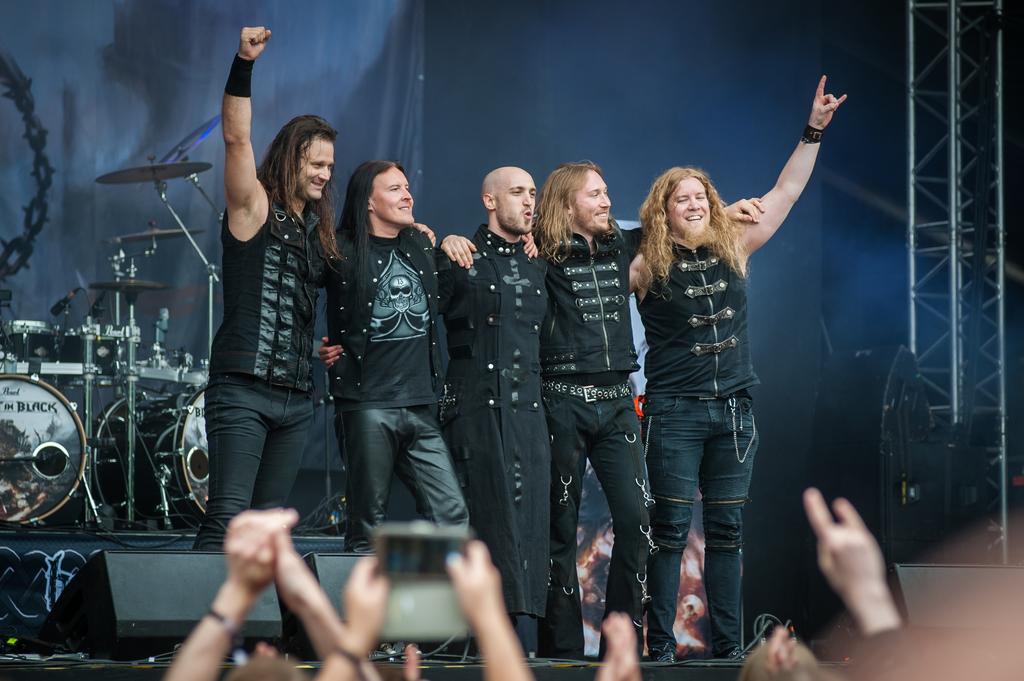}} & \parbox[t]{5cm}{$<$person$>$ is a $<$genre$>$ $<$genre$>$ band founded in Helsinki in 2015 by guitarist and $<$genre$>$ $<$person$>$ $<$person$>$. The $<$genre$>$ influences of the band ... \\} & \\
    & & Ground Truth & hard rock, metal, power metal, power pop, rock, synthpop \\ 
    & & SBNet (Multimodal) & \textcolor{red}{black metal}, \textcolor{red}{death metal}, \textcolor{red}{folk}, \textcolor{red}{folk metal}, \textcolor{red}{gothic metal}, \textcolor{blue}{hard rock}, \textcolor{red}{melodic death metal}, \textcolor{red}{melodic metal}, \textcolor{red}{melodic power metal}, \textcolor{blue}{metal}, \textcolor{blue}{power metal}, \textcolor{red}{progressive metal}, \textcolor{blue}{rock}, \textcolor{red}{speed metal}, \textcolor{red}{symphonic metal}, \textcolor{red}{symphonic power metal}, \textcolor{red}{thrash metal}, \textcolor{red}{viking metal} \\ 
    & & CLIP (Text-Only) & \textcolor{red}{black metal}, \textcolor{red}{death metal}, \textcolor{red}{folk metal}, \textcolor{red}{gothic metal}, \textcolor{blue}{hard rock}, \textcolor{red}{melodic death metal}, \textcolor{red}{melodic metal}, \textcolor{blue}{metal}, \textcolor{blue}{power metal}, \textcolor{red}{progressive metal}, \textcolor{blue}{rock}, \textcolor{red}{symphonic metal} \\ 
    & & CLIP (Image-Only) & \textcolor{red}{gothic metal}, \textcolor{red}{melodic metal}, \textcolor{blue}{metal}, \textcolor{blue}{power metal}, \textcolor{red}{progressive metal}, \textcolor{red}{symphonic metal} \\ 
    \midrule 
    \multirow{5}{*}{\textbf{Dannii Minogue}} & \multirow{5}{*}{\includegraphics[width=2cm,valign=c]{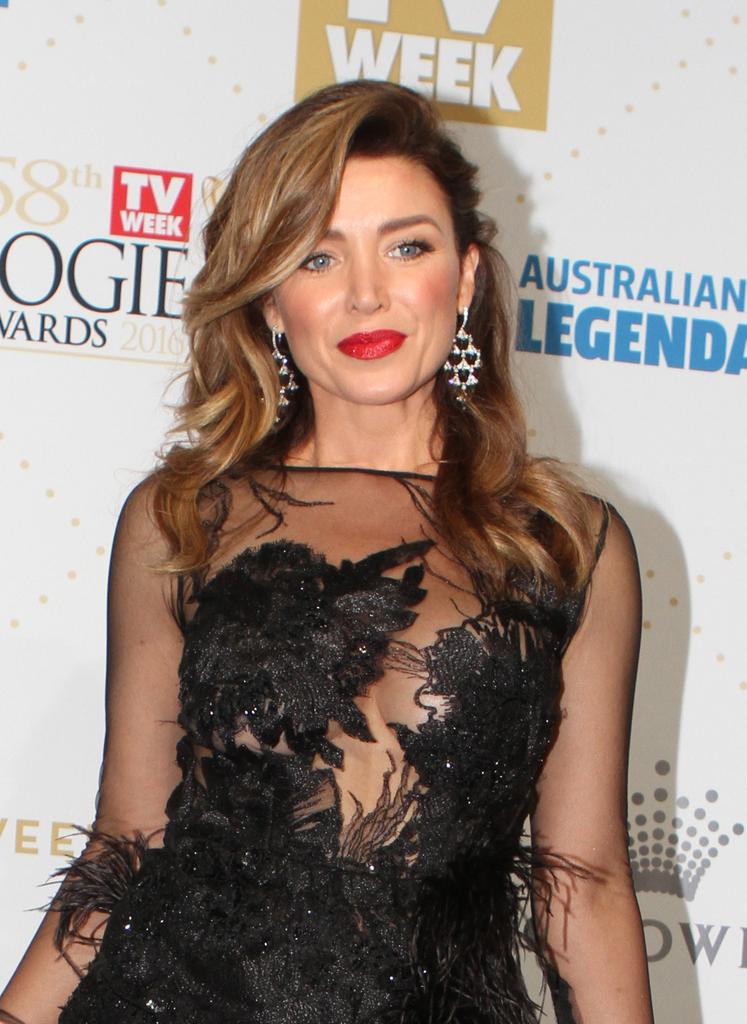}} & \parbox[t]{5cm}{$<$person$>$ $<$person$>$ (; born 20 October 1971) is an Australian $<$genre$>$, television personality, and actress. She first gained recognition .. \\} & \\
    & & Ground Truth & classic rock, dance pop, disco, electro, electronica, electropop, europop, freestyle, hi nrg, house, new age, nu disco, pop, pop dance, synthpop, trance, uk pop, world \\ 
    & & SBNet (Multimodal) & \textcolor{blue}{pop, dance pop, classic rock, synthpop} \\ 
    & & CLIP (Text-Only) & \textcolor{red}{alternative rock}, \textcolor{blue}{classic rock}, \textcolor{red}{country}, \textcolor{blue}{dance pop}, \textcolor{red}{easy listening}, \textcolor{blue}{pop}, \textcolor{red}{pop rock}, \textcolor{red}{rock}, \textcolor{red}{singer songwriter}, \textcolor{red}{soft rock}, \textcolor{red}{soul}, \textcolor{red}{soundtrack}, \textcolor{red}{teen pop} \\ 
    & & CLIP (Image-Only) & \textcolor{red}{easy listening}, \textcolor{red}{folk}, \textcolor{red}{jazz}, \textcolor{red}{lounge}, \textcolor{blue}{pop}, \textcolor{red}{rock}, \textcolor{red}{singer songwriter}, \textcolor{red}{soul}, \textcolor{red}{soundtrack} \\ 
\bottomrule 
\end{tabular} 
\end{table*}

\section{Dataset Curation}

    \ours{} builds upon the Music4All-Onion dataset~\cite{moscati2022music4all_onion}. We selected this dataset as a starting point since it already provides several multimodal representations for a large catalog of music \textit{tracks}, alongside user-item interactions, often missing in other multimodal music datasets. However, this dataset has the limitation that visual modalities are extracted from music video clips, and textual modalities from track lyrics. These might therefore not capture semantic similarities between music \textit{albums} or \textit{artists}. We therefore enrich the Music4All-Onion dataset with \textit{textual descriptions} and \textit{images} of the artists and albums of the tracks in the catalog. We would like to emphasize we do \textit{not} extract textual and image descriptors of albums and artists by aggregating representations of individual tracks. We instead leverage Music4All-Onion as a catalog source which allows us to identify extensive sets of artists and albums. 
    
    We extract metadata on artists and albums using the Last.fm API.\footnote{\url{https://www.last.fm/api}} 
    We first leverage the API to acquire the MusicBrainz Identifiers (MBIDs) of the music tracks in the Music4All-Onion dataset. These MBIDs are then used to query Last.fm and MusicBrainz\footnote{\url{https://python-musicbrainzngs.readthedocs.io/en/v0.7.1/api/}} for the metadata associated with the artist and album of the track
    . For artists, the metadata consists of name (e.g., `The Beatles' or `Bonnie Tyler'), sorted name (e.g., `Beatles, The' or `Tyler, Bonnie'), type (e.g., `Person', `Group'), gender, country of origin, life span, user-generated tags, and a list of relevant URLs related to the artist, such as fan pages, images, and Wikipedia pages.\footnote{\url{https://musicbrainz.org/doc/Style/Relationships/URLs}} Out of these, we use the URL to Wikidata to download the image of the artist using the Wikidata API (method `action=wbgetentities') and the Wikimedia Commons API (methods `action=query' and `prop=imageinfo`) to retrieve the images filenames and obtain the URL for download.

    For albums, the metadata consists of album name, performing artist, release date, number of distinct listeners, number of listens, and user-generated tags. We further obtain the album track listings, which provide track name, duration, and artist information, through Last.fm. For album cover images we use the Last.fm API (method `album.getInfo') to obtain the image URLs. If an album cover is not available via Last.fm, we use the same procedure described for the artists, relying on Wikimedia Commons
    . 
   
    For the textual descriptors of both artists and albums, from the list of MusicBrainz URL relations, we select the URL to the English Wikipedia page. If the URL to the Wikipedia page is not directly available, we use the URL to Wikidata to query the Wikidata API (methods `action=wbgetentities', `props=sitelinks') for the entity's Wikidata ID. This ID is then used to retrieve the title of the corresponding English Wikipedia page. With the Wikipedia page title identified, we then fetch the introductory paragraph using the Wikipedia API.

    For each entity, all collected metadata, textual descriptors and tags from both MusicBrainz and Last.fm are stored in individual JSON files named according to their MBID (`$<$MBID$>$.json'). Similarly, the corresponding downloaded images are stored as `$<$MBID$>$\_$<$index$>$.jpg', where `$<$index$>$' is an integer distinguishing the multiple images of a same entity. 

    We exclude albums and artists with incomplete information, restricting our dataset to those for which genres, textual descriptions, and images are all available.

    For our genre classification experiments, we partition the entities (i.e., the artists or the albums) separately into training and test sets. The initial split is performed using iterative stratification based on genre distributions, thereby ensuring that the training and test sets maintain a similar representation of genres. This process assigns approximately $80\%$ of the entities to the training set and the remaining $20\%$ to the test set. The resulting training and test splits are defined by lists of MBIDs stored in separate JSON files, one for artists and one for albums.

    To simulate missing modality scenarios, we generate specific modality splits for the test set. For various availability percentages $p\%$ ($10\%$, $30\%$, $50\%$, $70\%$, $90\%$, $100\%$), we define subsets of the test set.
    
    Specifically, for a given percentage $p\%$, a random subset containing $p\%$ of the total test entities is selected. For these entities, the modality is considered available. Consequently, for the remaining $(1-p)\%$ of the test entities, that modality is treated as missing during evaluation. These percentage-based subsets are constructed to be nested, meaning that the set of entities for which a modality is available at $p_1\%$ is a subset of those available at $p_2\%$ if $p_1 < p_2$. This ensures that, as we increase the modality availability percentage, we are only adding more entities with that modality, not changing the previously included ones. The $100\%$ split represents the full test set where the modality is considered available for all test items.

\begin{table}[t]
\centering
\caption{Performance comparison on \ours{} dataset under unimodal and multimodal configurations. F$1$ score are sample-averaged. Higher F$1$ score indicate a better performance ($\uparrow$).}

\resizebox{0.45\textwidth}{!}{%
    \begin{tabular}{@{}lrrc@{}}
    \toprule
    \textbf{Dataset} & \textbf{Config.} & \textbf{Method }& \textbf{F1 Samples ($\uparrow$)} \\
    \midrule
    \multirow{5}{*}{\ours{} (Albums)} & Image-only & CLIP~\cite{radford2021learning} & 31.3 \\
                                       & Text-only  & CLIP~\cite{radford2021learning} & 25.4 \\
    \cline{2-4}
                                       &  \multirow3{*}{Multimodal}
                                       & MMBT~\cite{kiela2019supervised} & 14.5 \\
                                       & & ViLT~\cite{kim2021vilt} & 17.4 \\
                                       & & SBNet~\cite{saeed2023single} & 33.1 \\
    \midrule\midrule
    \multirow{5}{*}{\ours{} (Artists)} & Image-only & CLIP~\cite{radford2021learning} & 26.0 \\
     & Text-only & CLIP~\cite{radford2021learning} & 23.9 \\
   \cline{2-4}
    & \multirow{3}{*}{Multimodal}
      & ViLT~\cite{kim2021vilt}         & 16.6 \\
    &  & MMBT~\cite{kiela2019supervised} & 16.8 \\
    & & SBNet~\cite{saeed2023single}    & 29.4 \\
    \bottomrule
    \end{tabular}
    \label{tab:music4all_performance}
    }%
\end{table}

\begin{table}[t]
\centering
\caption{Performance comparison on MM-IMDb dataset. F$1$ score are sample-averaged.  Higher F$1$ score indicate a better performance ($\uparrow$).}

    \begin{tabular}{@{}lrrc@{}}
    \toprule
    \textbf{Dataset} & \textbf{Config.} & \textbf{Method} & \textbf{F1 Samples ($\uparrow$)} \\
    \midrule
    \multirow{10}{*}{MM-IMDb} &\multirow{2}{*}{Text-only}
     & CLIP~\cite{radford2021learning}   & 58.3 \\
    & & MFAS~\cite{perez2019mfas}      & 60.6 \\
    \cline{2-4}
    & \multirow{2}{*}{Visual-only}
    & MFAS~\cite{perez2019mfas}      & 48.4 \\
    & & CLIP~\cite{radford2021learning}   & 60.2 \\
     \cline{2-4}
    & \multirow{6}{*}{Multimodal}
    &   MFAS~\cite{perez2019mfas}                & 62.5 \\
    & & GMU~\cite{arevalogated2017}              & 63.0 \\
    & & MMBT~\cite{kiela2019supervised}          & 63.6 \\
    & & CentralNet~\cite{vielzeuf2018centralnet} & 63.9  \\
    & & ViLT~\cite{kim2021vilt}                  & 64.6   \\
    & & SBNet~\cite{saeed2023single}             & 65.3    \\

    \bottomrule
\end{tabular}
\label{tab:sota_imdb}
\end{table}

    \section{Experiments}  

    We benchmark our dataset on the task of multimodal music genre classification, leveraging both images and textual descriptors. Furthermore, in order to assess how well images and textual descriptors separately encode information on the genres, we consider scenarios in which one of the modalities is missing for some of the entities.

    \subsection{Data Preprocessing}
    The ground truth genre labels are derived from the metadata obtained via the Last.fm API. This metadata includes a list of user-generated tags attached to a weight between 1 and 100, corresponding to the frequency of its occurrence for the album or artist under consideration.
    
    To curate a consistent set of genre labels from these potentially noisy and diverse user tags, we perform a two-step refinement process. First, we normalize each candidate tag from Last.fm. This normalization involves converting the tag to lowercase and removing leading/trailing whitespace to ensure consistent matching. Second, these normalized tags are filtered against a reference list of music genres and styles that we compile based on the information available on Wikipedia.\footnote{Based on \url{https://en.wikipedia.org/wiki/List_of_music_genres_and_styles}} A normalized Last.fm tag is accepted as a valid genre label only if it exactly matches an entry in our reference list.
    
    To reduce potential information leakage from our textual descriptors to the genre labels, we apply two masking strategies. First, recognizing that Wikipedia summaries may explicitly mention genres, we mask terms corresponding to genre classes from our reference genre list that occurr within these texts by replacing them with a $<$genre$>$ placeholder. Second, since artist or album names can themselves be strong indicators of certain genres, and to avoid giving a strong advantage to pre-trained text encoders, we employ Named Entity Recognition to identify and anonymize mentions of artist and album names within the textual descriptions, replacing them with the $<$person$>$ placeholder.
    
    We standardized visual content by resizing images exceeding 1024×1024 pixels to fit within these dimensions while preserving aspect ratios. As described above, we applied a stratified sampling approach based on genre distributions to ensure representative class balances between training and evaluation sets.

    \subsection{Baselines}
        Following prior work~\cite{pramanick2021,capAlign}, we use CLIP~\cite{radford2021learning} as baseline \textit{unimodal} model to evaluate image- and text-only performances. Specifically, we utilize the pre-trained openai/clip-vit-large-patch14 model\footnote{\url{https://huggingface.co/openai/clip-vit-large-patch14}} to extract fixed embeddings for images and text.
        The pre-computed CLIP embeddings are first 
        normalized, and then processed by a Multi-Layer Perceptron with a final linear classification layer with sigmoid activation.
        
        In line with several recent works
        ~\cite{pramanick2021,capAlign,arevalogated2017}, we use the following \textit{multimodal} models. 
        
        \noindent \textbf{Multimodal BiTransformer (MMBT):} This architecture  integrates a pre-trained BERT encoder for textual representation with a ResNet-$152$ backbone for visual feature extraction, followed by a cross-modal fusion mechanism~\cite{kiela2019supervised}.
        
        \noindent \textbf{Vision-and-Language Transformer (ViLT):} Utilizing a pre-trained vision-language transformer, this approach implements joint encoding of visual and textual modalities through a unified transformer architecture with learned cross-attention mechanisms~\cite{kim2021vilt}.
        
        \noindent \textbf{Single-branch Network (SBNet):} This  architecture diverges from traditional dual-branch approaches by implementing a unified modality-invariant branch that processes both visual and textual representations within the same network~\cite{saeed2023single}.

        \subsubsection*{Baseline Performance Validation}
        To ensure that the selected models are effective for multimodal genre classification,  
        we evaluate them not only on our newly curated dataset, but also on  MM-IMDb~\cite{arevalogated2017}, a well-established dataset for benchmarking genre classification models. On this dataset, the selected baselines perform comparably to state-of-the-art methods such as MFAS~\cite{perez2019mfas}, CentralNet~\cite{vielzeuf2018centralnet}, and GMU~\cite{arevalogated2017}, as reported in Table~\ref{tab:sota_imdb}.
        This indicates that the selected baselines are meaningful to benchmark our dataset. 

        \subsection{Implementation Details}
        We train MMBT, ViLT, and SBNet with the same hyperparameters reported in prior works~\cite{kiela2019supervised,kim2021vilt,saeed2023single}. For all models, the final layer outputs logits for each genre. To perform multi-label genre classification, a sigmoid activation function is applied element-wise to these logits. A genre is then predicted as present for an item if its corresponding sigmoid output exceeds a threshold of 0.5. For model optimization, we employ binary cross-entropy loss, incorporating class-frequency-based weighting to address label imbalance. The optimization process implements early stopping based on validation F$1$ scores to prevent overfitting. We evaluate the performance on held-out test sets using F$1$ score, a standard multi-label classification metric.

        Moreover, we systematically assess robustness in missing-modality settings.
        Following prior work~\cite{ma2022multimodal,liaqat2025chameleon,saeed2023single} and as described above, we conduct experiments across varying proportions of available textual and visual information. For each modality and percentage $p\%$, we randomly select $(1 - p)\%$ of the catalog entities in the test set. For these items, the selected modality is considered as missing during the evaluation time. This approach enables comparison of the robustness of different models to missing modality scenarios. For the \ours{} dataset, this controlled ablation approach allows us to evaluate which modality (text or image) is semantically more relevant for music genre classification of albums and artists.

        \subsection{Results}
        To validate the effectiveness of the selected baseline methods, we evaluate them on the benchmark MM-IMDb dataset.
        As shown in Table~\ref{tab:sota_imdb}, all multimodal approaches consistently outperform unimodal configurations, with SBNet achieving the highest F$1$ score ($65.3\%$).  The other two selected baselines (MMBT and ViLT) achieve a lower F$1$ score, but are still competitive with other state-of-the-art methods. These results confirm the validity of our baseline selection. 

        Despite this validation of the selection of our baseline models, it is important to remark that a direct comparison between the MM-IMDb (movie) and Music4All A+A (music) datasets has to be carried out considering the different characteristics of the two datasets. Beyond the domain difference, \ours{} contains substantially more genre labels (659 for artists and 737 for albums) compared to MM-IMDb's 26 genres, and the genres themselves are inherently different between the two domains. This difference in label space complexity and domain-specific genre definitions may contribute to the performance variations observed between datasets. Nevertheless, the evaluation on MM-IMDb serves to validate that our selected baseline methods are capable of achieving competitive performance on established benchmark datasets.
        
        Fig.~\ref{tab:qualitative_sbnet_clip_artist_variants} shows qualitative examples of model predictions on artists, showing the two modalities (image and text), the masking technique, as well as the ground-truth and predicted genres.

        Table~\ref{tab:music4all_performance} compares the performance of unimodal and multimodal methods on the \ours{} dataset, evaluated using sample-averaged F$1$ score and considering both textual and visual modalities as fully available ($100\%$). The results are presented separately for the Album and Artist subsets. For both subsets, CLIP achieves competitive performance in unimodal settings (image-only and text-only). 
        On both subsets, we observe that CLIP performs better when leveraging images compared to text, suggesting that images of artists and album covers are more representative of music genres, compared to textual descriptions. This is in contrast with the results on MM-IMDb reported in Table~\ref{tab:sota_imdb}, where text-based CLIP and text-based MFAS outperform the image-based variants. This is a first indication that the relevance of textual and image modalities for genre classification depends on the domain considered. Multimodal methods exhibit varied performance. Notably, SBNet outperforms other methods, achieving the highest F$1$ score ($33.1$\% for Albums, $29.4$\% for Artists). 
        In contrast, MMBT and ViLT reach a lower score than all other models, including unimodal ones. The fact that these models achieve performance competitive with state-of-the-art methods on MM-IMDb but not on our datasets suggests that multimodal models might face limitations in their adaptation depending on the domain. 
        These results therefore highlight the importance of domain-dependent method selection in multimodal learning, and suggest that SBNet is more versatile. Since SBNet achieves the best performance across all datasets when all modalities are available, we base our analysis of missing-modality scenarios on this model. Table~\ref{tab:f1_score} presents the performance of SBNet on MM-IMDb and our dataset under varying missing-modality scenarios. The first two columns indicate the percentages of catalog items for which the corresponding modality is available. A percentage of $100\%$ indicates that the corresponding modality is available for all items. On all datasets, the performance of SBNet is highest when both modalities are available for all items, and deteriorates whenever either image or text modality is partially missing. For our dataset, the performance decrease is particularly remarkable for the Albums subset when the image modality is missing ($33.1\%$ to $13.0\%$), as compared to missing texts ($33.1\%$ to $30.5\%$), indicating that album covers are highly informative in assessing the music genre. Notably, when text modality is missing for a large portion of the catalog, e.g., when only available for $10$\% of the catalog, the multimodal performance is even worse than the unimodal one. This deteriorated performance indicates that multimodal models might be ineffective for real-world scenarios of missing modality.

\begin{table}[t]
  \centering
  \caption{Test F$1$ scores for SBNet with varying modality presence on MM-IMDb and our dataset.}
  \label{tab:f1_score}
  \begin{tabular}{@{}cc | cHH || cHH | cHH@{}}
  \toprule
  Text \% & Image \% & MM-IMDb &  &  & Albums &  & & Artists &  &    \\
  \midrule
    100 & 100 & 65.3 & 58.0 & 63.6 & 33.1 & 17.4 & 14.6 & 29.4 & 16.6 & 15.4 \\
    \midrule
    100 & 90 & 62.5 & 54.5 & 64.0 & 31.1 & 15.3 & 14.6 & 27.7 & 15.2 & 15.4 \\
    100 & 70 & 56.9 & 49.5 & 52.7 & 28.9 & 14.6 & 14.6 & 26.0 & 15.1 & 15.4 \\
    100 & 50 & 50.8 & 47.1 & 47.7 & 26.4 & 14.6 & 14.6 & 24.3 & 13.4 & 15.4 \\
    100 & 30 & 45.2 & 42.1 & 31.3 & 25.2 & 14.6 & 14.6 & 22.8 & 10.3 & 15.4 \\
    100 & 10 & 39.8 & 33.2 & 27.0 & 13.0 & 14.6 & 14.6 & 21.6 & 10.3 & 15.4 \\
   \midrule
    90 & 100 & 63.2 & 57.7 & 64.1 & 31.3 & 15.2 & 14.6 & 25.3 & 14.8 & 15.4 \\
    70 & 100 & 59.3 & 56.9 & 61.7 & 32.8 & 14.6 & 14.6 & 25.9 & 10.3 & 15.4 \\
    50 & 100 & 55.6 & 55.4 & 61.6 & 30.8 & 14.6 & 14.6 & 24.0 & 10.3 & 15.4 \\
    30 & 100 & 51.8 & 54.7 & 61.4 & 29.1 & 14.6 & 14.6 & 22.6 & 98.5 & 15.4 \\
    10 & 100 & 48.0 & 52.3 & 61.5 & 30.5 & 14.6 & 14.6 & 21.5 & 10.8 & 15.4 \\
    \bottomrule
  \end{tabular}
\end{table}

\section{Conclusions}
    In this paper, we introduced \ours, a multimodal dataset of music \textit{artists} and \textit{albums}, comprising genres, images, and textual descriptors. 
    The dataset covers $6,741$ artists and $19,511$ albums, providing a comprehensive resource for multimodal music information retrieval tasks.
    We established baseline results for album and artist genre classification on \ours{} using both unimodal and multimodal configurations. The results show that for unimodal models, images lead to more accurate genre classification than textual descriptors, and that leveraging both modalities leads to better results.  
    Our analysis also showed that several multimodal learning models that are effective for genre classification in the movie domain are not effective in the music domain, demonstrating a strong domain dependence of multimodal models. To address this limitation and to further investigate the under-performance of several models on the \ours{} dataset, we propose leveraging the dataset to organize an ACM MM challenge, advancing the development of robust multimodal learning methods capable of handling incomplete data and across more than one domain.
    
    This work has several limitations. First and foremost, \ours{} is suitable for many MIR tasks beyond genre classification, such as auto-tagging and music recommendation, which were not considered in our benchmark experiments. Furthermore, our experiments relied only on the data provided by \ours{} dataset. Since \ours{} is built on top of the Music4All-Onion dataset, more extensive experiments could include the use of more modalities, including audio and user--item interaction data. Finally, since the genres in \ours{} are extracted from user-generated tags from Last.fm, they are also associated with weights; this would allow to reformulate the task of genre classification as regression, i.e., to predict the relative importance of a genre for a specific artist or album with respect to the others. We leave these extensions of our work for future research. 
\section*{Acknowledgment}
    This research was funded in whole or in part by the Austrian Science Fund (FWF) \url{https://doi.org/10.55776/P33526}, \url{https://doi.org/10.55776/DFH23}, \url{https://doi.org/10.55776/COE12}, \url{https://doi.org/10.55776/P36413}. 
\balance
\bibliographystyle{IEEEtran}
\bibliography{samples/IEEEabrv,samples/2025_CBMI_bib}

\end{document}